\input psfig.sty
\documentstyle{cupconf}

\ifoldfss
\else
  \ifnfssone
    \newmathalphabet{\mathit}
      \addtoversion{normal}{\mathit}{cmr}{m}{it}
      \addtoversion{bold}{\mathit}{cmr}{bx}{it}
    \newmathalphabet{\mathcal}
      \addtoversion{normal}{\mathcal}{cmsy}{m}{n}
    \else
    \ifnfsstwo
    \fi
  \fi  
\fi

\def\hexnumber#1{\ifcase#1 0\or1\or2\or3\or4\or5\or6\or7\or8\or9\or
 A\or B\or C\or D\or E\or F\fi }

\makeatletter
\ifx\CUP@mtlplain@loaded\undefined
\else
\fi
\makeatother
 \makeatletter
 \ifx\CUP@mtlplain@loaded\undefined
   \font\tenbmi=cmmib10 at 10pt
   \font\sevenbmi=cmmib10 at 7pt
   \font\fivebmi=cmmib10 at 5pt

   \newfam\bmifam
   \textfont\bmifam=\tenbmi
   \scriptfont\bmifam=\sevenbmi
   \scriptscriptfont\bmifam=\fivebmi
   
 \fi
 \makeatother
\ifnfsstwo

\fi
\ifnfssone

\fi
\ifoldfss

\fi

\mathchardef\varLambda="0103

\makeatletter
\ifx\CUP@mtlplain@loaded\undefined
\else
\fi
\makeatother
\makeatletter
\ifx\CUP@mtlplain@loaded\undefined
  \font\tenbms=cmbsy10
  \font\sevenbms=cmbsy10 at 7pt
  \font\fivebms=cmbsy10 at 5pt
  \newfam\bmsfam
  \textfont\bmsfam=\tenbms
  \scriptfont\bmsfam=\sevenbms
  \scriptscriptfont\bmsfam=\fivebms

  \edef\bsy@{\hexnumber\bmsfam}
  \mathchardef\bnabla="0\bsy@72
\fi
\makeatother
\title[HI Absorption toward NGC 3894]{Global VLBI Observations of HI Absorption toward NGC 3894}

\author[Peck \& Taylor ]{%
A.\ns B.\ns P\ls E\ls C\ls K$^1$\ns \and \ns G.\ns B.\ns T\ls A\ls Y\ls L\ls O\ls R$^2$}

\affiliation{$^1$MPIfR, Auf dem H\"ugel 69, D-53121 Bonn, Germany\\[\affilskip]
$^2$NRAO, P.O. Box O, Socorro, NM 87801, USA} 

\begin{document}
\ifnfssone
\else
  \ifnfsstwo
  \else
    \ifoldfss
      \let\mathcal\cal
      \let\mathrm\rm
      \let\mathsf\sf
    \fi
  \fi
\fi

\maketitle

\begin{abstract}

One of the most important problems in the study of AGN is
understanding the detailed geometry, physics, and evolution of the
central engines and their environments.  The leading models involve an
accretion disk and torus around a central black hole.  Much of this
torus should be comprised of atomic gas, detectable in absorption
toward the bright inner radio jets.  In the last few years, a number
of compact symmetric radio sources have been found to exhibit
H\kern0.1em{\sc i}\ absorption, at or near the systemic velocity,
toward the central parsecs.  Understanding the kinematics of the
H\kern0.1em{\sc i}\ detected toward the central parsecs of these
sources will provide an important test of this model and of unified
schemes for AGN.

We present results of Global VLBI Network observations at 1.4 GHz
toward the active nucleus of the nearby elliptical galaxy NGC 3894
(a.k.a. 1146+596, $z$=0.01068).  The center of activity in this source
and the orientation of the jets with respect to our line of sight have
been determined using VLBI studies of the proper motions of jet
components.  The 21 cm atomic hydrogen line is seen in absorption
slightly redshifted with respect to the systemic velocity toward the
core, jet, and counterjet of this source.  The structure of the
H\kern0.1em{\sc i}\ in this source is complicated.  We find several
distinct components present along the lines of sight to the
approaching and receding jets, making interpretation challenging.

\end{abstract}

\section{Introduction}
NGC 3894 is an elliptical galaxy at a redshift of $z$=0.01068
(Karachentsev 1980).  VLBI studies of the proper motions in the jet
components at 8 GHz by Taylor, Wrobel \& Vermeulen (1998) indicate
that the jets are probably oriented at about 50$^{\circ}$~ to the line
of sight, with the northwest side being the approaching jet.  The core
is indicated with an asterisk in Figure 1.  The H\kern0.1em{\sc i}\
absorption in this source was originally detected by van Gorkom et al
(1989), and followup observations were made with the VLBA and phased
VLA in 1994.  The 1994 observations had an angular resolution of 9 mas
and an rms noise of 1.5 mJy/beam/channel (Peck \& Taylor 1998). Here
we present the preliminary results of more recent Global VLBI
observations made in Nov. 1998.  In these observations, the angular
resolution is $\sim$6$\times$4 mas, corresponding to $\sim$1.5 pc
(assuming H$_0$=75 km s$^{-1}$~Mpc$^{-1}$).  The rms noise is $\sim$0.3
mJy/beam/channel.

 \begin{figure}[!h]
 \centerline{\psfig{figure=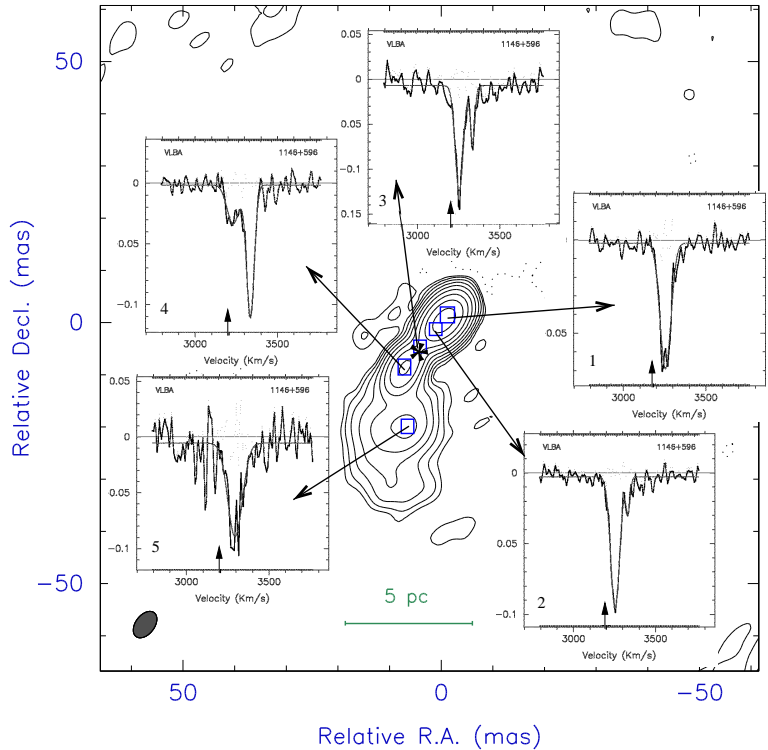,height=4.5in}}
 \caption{Integrated H\kern0.1em{\sc i}\ absorption profiles toward NGC~3894.  The systemic velocity is indicated by an arrow in each panel.  Parameters determined from Gaussian fits to the profiles are shown in Table 1.}
 \end{figure}

\renewcommand{\baselinestretch}{1.2}
\begin{table}
\begin{center}
\begin{tabular}{ccccccc}
Profile&Component & Amplitude &Central Vel. &FWHM&$\tau$&N$_{\rm HI}$ \\
& &(mJy) &(km s$^{-1}$) &(km s$^{-1}$) && (10$^{20}$~cm$^{-2}$) \\
1&a&8.8$\pm$0.2&3255.0$\pm$1.0&78.4$\pm$2.4&0.068$\pm$0.003&9.7\\
2&a&9.9$\pm$0.2&3253.5$\pm$0.7&64.0$\pm$1.7&0.094$\pm$0.002&11.0\\
&b&2.7$\pm$0.3&3332.1$\pm$1.7&29.6$\pm$4.1&0.025$\pm$0.003&1.3\\
3&a&4.2$\pm$0.2&3253.0$\pm$1.2&57.8$\pm$2.9&0.118$\pm$0.005&12.4\\
&b&2.2$\pm$0.2&3334.9$\pm$1.6&32.0$\pm$3.8&0.066$\pm$0.006&3.9\\
4&a&1.6$\pm$0.1&3230.8$\pm$3.6&87.6$\pm$9.5&0.032$\pm$0.002&5.1\\
&b&5.3$\pm$0.2&3335.7$\pm$0.9&55.5$\pm$2.1&0.109$\pm$0.003&11.0\\
5&a&0.8$\pm$0.2&3294.4$\pm$3.5&104.6$\pm$8.7&0.084$\pm$0.006&16.0\\
\end{tabular}
\caption{Parameters Determined from Gaussian Fits to the Absorption Profiles}
\end{center}
\end{table}
\firstsection 

\section{Discussion}

Figure 1 shows the integrated H\kern0.1em{\sc i}\ absorption profiles
toward NGC 3894. Each profile is integrated over 9 pixels in the data,
encompassing an area slightly smaller than the beam.  The systemic
velocity is indicated by the arrows in each panel.  The velocity
centroid of the absorption in each panel is within 150 km s$^{-1}$ of
the systemic velocity.  There are two distinct velocity components
toward the central regions of the source, shown in Profiles 2, 3 and
4.  In Profiles 1 and 5, the lines are blended and only one Gaussian
component could be fit.  The parameters determined from Gaussian fits
to these components are shown in Table 1.  The column density, N$_{\rm
HI}$, was estimated assuming a spin temperature of 100 K and a covering
factor of 1.  If the gas is associated with a circumnuclear torus,
T$_{\rm spin}$ will probably be much higher.  For each panel, the lower
velocity, broader component is referred to as Component a in the
table.

In addition to the integrated profiles shown in Figure 1, two Gaussian
components have been fit to the data at each pixel across the source.
Pixels in which the signal to noise ratio was less than 2 have been
blanked.  The results of these fits are shown in Figures 2 and 3.  The
horizontal lines seen in these images immediately to the east of the
secondary continuum peak are due to the algorithm used in the Gaussian
fitting program.  An initial guess is made for each fit based on the
last recorded good fit, resulting in different fits in the presence of
blended lines which depend on the direction of rastering in each row.
This problem will soon be overcome by making the fit interactively at
each pixel.  In the meantime, the general trends can still be seen
over the rest of the continuum source.

Figure 2 shows the optical depth and FWHM linewidth distributions of
the broad H\kern0.1em{\sc i}\ line.  Although the optical depth is
significantly higher toward the approaching jet than the receding jet,
the sharp gradient in the region of the core demonstrates that
optical depth is sensitive to the location of the nucleus, which
would not be the case with a distant foreground cloud.  The linewidth
peaks south of the core (FWHM$\sim$275 km s$^{-1}$), and this peak
corresponds to a region of rising optical depth which would be
consistent with a line of sight through a long pathlength in a thick
torus.  Further analysis is required to determine whether the FWHM of
the line is as high to the NNE of this region, where the rastering
problem has yet to be eliminated.  If the FWHM is high in this region,
this could be evidence of the central few parsecs of a torus which, in
projection, appears to be centered about 1 pc from the core.  The
lower optical depth in this region then might be attributable to a
higher spin temperature in close proximity to the powerful central
x-ray source.

Figure 3 shows the optical depth and FWHM linewidth distributions of
the narrow line.  The FWHM indicates a region of higher velocity
dispersion (FWHM$\sim$60 km s$^{-1}$) toward the inner receding jet.
The optical depth seems to rise toward the core as well.  The
variations in optical depth further along the receding jet, however,
are not consistent with a simple torus model.  The absorption features
centered on this velocity appear substantially narrower than what we
have come to expect from a torus, based on sources like 1946+708
(Peck, Taylor \& Conway 1999) and PKS 2322-123 (Taylor et al 1999).
The higher velocity with respect to the systemic of this narrow line
indicate that the absorption might be due to the presence of
disorganized inward streaming gas or multiple infalling clouds along
the line of sight.

 \begin{figure}[!h]
 \centerline{\psfig{figure=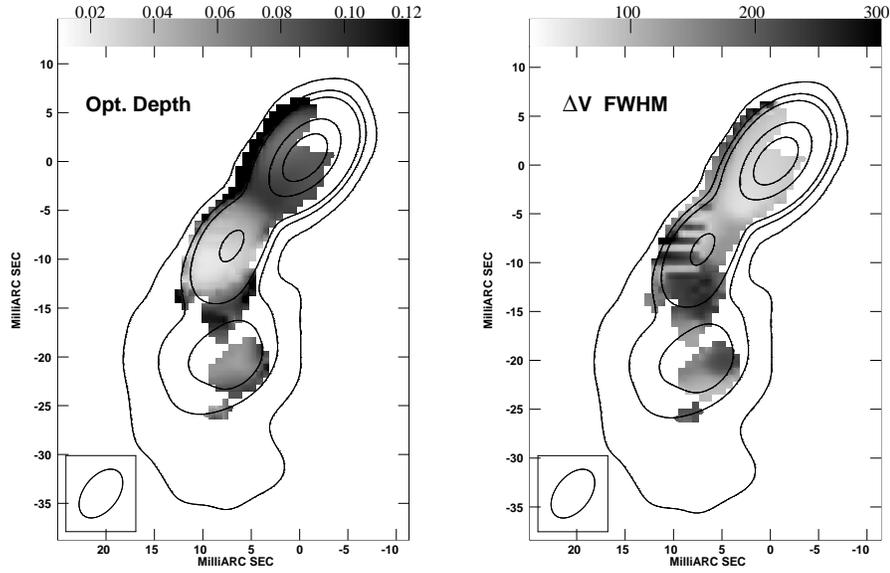,height=9cm}}
 \caption{Distribution of optical depths and linewidths of the
 broader H\kern0.1em{\sc i}\ component toward NGC~3894.  The beam is
 shown in the lower left.}  \end{figure} 

\begin{figure}[!h]
 \centerline{\psfig{figure=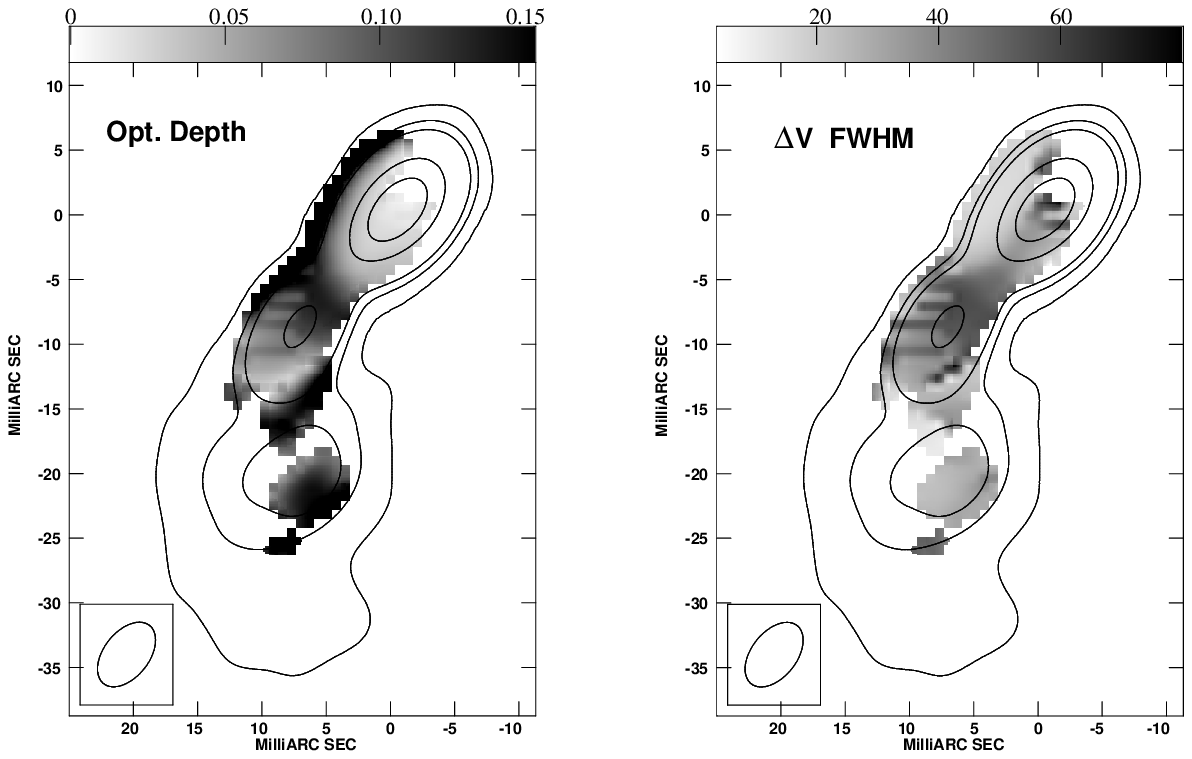,height=9cm}}
 \caption{Distribution of optical depths and linewidths of the narrower
 H\kern0.1em{\sc i}\ component toward NGC~3894.  The beam is shown in
 the lower left.}  \end{figure}

\end{document}